# Social media uptake of scientific journals: A comparison between X and WeChat

Ting Cong[1], Er-Te Zheng[2], Zekun Han[3], Zhichao Fang[3,4*], Rodrigo Costas[4,5]

* Corresponding author: z.fang@cwts.leidenuniv.nl

1. College of Publishing, University of Shanghai for Science and Technology, Shanghai, China.

2. School of Information, Journalism and Communication, The University of Sheffield, Sheffield, UK.

3. School of Information Resource Management, Renmin University of China, Beijing, China.

4. Centre for Science and Technology Studies (CWTS), Leiden University, Leiden, The Netherlands.

5. DSI-NRF Centre of Excellence in Scientometrics and Science, Technology and Innovation Policy, Stellenbosch University, Stellenbosch, South Africa.

**Abstract**: This study examines the social media uptake of scientific journals on two different platforms – X and WeChat – by comparing the adoption of X among journals indexed in the Science Citation Index-Expanded (SCIE) with the adoption of WeChat among journals indexed in the Chinese Science Citation Database (CSCD). The findings reveal substantial differences in platform adoption and user engagement, shaped by local contexts. While only 22.7% of SCIE journals maintain an X account, 84.4% of CSCD journals have a WeChat official account. Journals in Life Sciences & Biomedicine lead in uptake on both platforms, whereas those in Technology and Physical Sciences show high WeChat uptake but comparatively lower presence on X. User engagement on both platforms is dominated by low-effort interactions rather than more conversational behaviors. Correlation analyses indicate weak-to-moderate relationships between bibliometric indicators and social media metrics, confirming that online engagement reflects a distinct dimension of journal impact, whether on an international or a local platform. These findings underscore the need for broader social media metric frameworks that incorporate locally dominant platforms, thereby offering a more comprehensive understanding of science communication practices across diverse social media and contexts.

**Keywords**: Altmetrics, social media metrics, science communication, scholarly publishing, research evaluation

## 1. Introduction

Over the past two decades, the rapid proliferation of social media has profoundly reshaped how science is broadly communicated, expanding both the pathways through which scientific information is disseminated and the diversity of audiences that can be reached [1, 2]. Multiple social media platforms now provide channels for immediate scientific communication and real-time interaction among users from varied backgrounds [3], thereby transcending the traditional boundaries of academic journals and conferences [4]. As scientific content circulates more widely online, opportunities arise to measure how science is shared, understood, and engaged with by broader audiences [5, 6]. This development has spurred the emergence of *social media metrics* [7], which focus on capturing and analyzing evidence of online attention related to academic content and actors.

Initially, social media metrics were primarily envisioned as a complement to traditional citation-based indicators, aiming to quantify forms of attention distinct from those represented by citations [8–10]. However, there is a growing consensus that social media metrics do more than simply measure the visibility of research outputs, but for example, also involve studying the identities and activities of users engaging with scientific content [7, 11]. To more thoroughly understand the contexts in which science is disseminated on social media, a substantial body of literature has emerged that explores how diverse stakeholders – both within and outside the academic community – adopt social media platforms to disseminate and share scientific content, and what motivates their engagement with these platforms.

*1.1. Social media uptake of various academic actors*

Given the public nature of many social media platforms, the dissemination of scientific content on these platforms was initially viewed as an indicator of public engagement with science [12, 13]. However, numerous studies that investigate the identities of social media users reveal that a considerable proportion of those who interact with scientific content are themselves affiliated with academia [11, 14–18]. For instance, in a large-scale study comprising approximately 2.5 million X (formerly Twitter)[1] users who tweeted about scientific papers, Zhang et al. [19] found that half of the users were associated with the academic community and that they were responsible for the majority of scientific tweets. Likewise, a manual coding analysis by Yu et al. [20] showed that about 49% of X users mentioning scientific papers were researchers, while a user survey by Mohammadi et al. [21] indicated that roughly 55% of individuals tweeting about scientific papers worked in academia. Taken together, these findings underscore that science-related discussions on social media reflect a mixed mode of participation, involving both academic and non-academic users.

Recognizing the critical role of academic users in online science communication, several studies have explored the social media uptake of various academic actors, including individual scholars [22–25], universities [26, 27], scientific journals [28–30], scholarly publishers [31, 32], and also bots [33]. The extent of social media uptake differs across various types of academic users and platforms, and is further modulated by demographic factors such as discipline, gender, and academic seniority [34–36]. These studies have not only advanced methods for identifying specific categories of academic users on social media but have also illuminated how particular user groups facilitate the dissemination of scientific information in online environments.

*1.2. Social media uptake of scientific journals across platforms*

As social media becomes increasingly integrated into academic workflows, scientific journals have also embraced these platforms as a means of engaging and retaining readers, as well as boosting global visibility [37, 38]. Previous research indicates that many journals now maintain a presence on social media, although the degree of uptake varies across disciplines and platforms. For instance, on X, the proportion of journals with an active presence ranges from 22.2% among those indexed in the *Science Citation Index-Expanded* (SCIE), to 27.6% in the *Arts and Humanities Citation Index* (AHCI), and up to 35.1% in the *Social Science Citation Index* (SSCI) [29]. Similarly, 24.2% of

---

[1] Twitter was officially rebranded as X in 2023. However, since the platform was still known as Twitter during our data collection period, we continue to use the established terminology – such as tweets, retweets, likes, replies, and quotes – throughout this study.

Urological journals maintain an X account [39]. In contrast, 65.3% of journals indexed in the *Chinese Social Science Citation Index* (CSSCI) have established a presence on WeChat [28]. Such disparities underscore the diverse contexts and strategic considerations influencing the social media adoption decisions of scientific journals.

Beyond disciplinary differences, studies have also compared the extent to which scientific journals utilize various social media platforms. X and Facebook are the two most frequently examined international social networks in science communication. For example, among 25 Medical journals analyzed by Boulos and Anderson [40], 20 had a presence on Facebook, whereas only 11 were on X. By contrast, Amir et al. [41] found that the X uptake of Dermatology journals exceeded their Facebook uptake (13.7% vs. 12.7%). In the case of social media uptake among both high-impact Library and Information Science journals and Communication journals, X was found to be the predominant platform, surpassing others such as Facebook, LinkedIn, Instagram, and YouTube [42]. Similarly, among 100 Web of Science-indexed journals with high impact factors, the percentage of journals on X (16%) was higher than those on Facebook (9%) [43]. However, in another analysis, while Facebook uptake exceeded X uptake for AHCI-indexed journals (14.2% vs. 9.0%), SCIE-indexed and SSCI-indexed journals exhibited a greater presence on X than on Facebook (9.7% vs. 7.7% and 7.7% vs. 7.2%, respectively) [30].

*1.3. Objectives of the study*

While previous research has significantly advanced the understanding of how international scientific journals utilize global social media platforms (e.g., X and Facebook), relatively little attention has been paid to how local journals adopt local social media platforms. Yu et al. [44] proposed a framework that classifies social media metric studies into four types, based on the geographic scope of both the social media sources and the scientific outputs being analyzed. Most existing work on the social media uptake of journals falls under the category of *international research discussed on international platforms*. A smaller body of work has examined *local research discussed on international platforms*, such as studies investigating the international social media uptake of Ibero-American scientific journals [45, 46]. By contrast, studies focusing on *local research discussed on local platforms* remain rare.

In this study, we address this gap by examining two platforms with distinct user bases and cultural contexts: X, an internationally recognized social media platform, and WeChat, a leading social media application in China. Specifically, we compare how journals indexed in the *Science Citation Index-Expanded* (SCIE) utilize X with how those indexed in the *Chinese Science Citation Database* (CSCD) utilize WeChat. Through this comparative lens, we aim to enhance our understanding of how scientific journals adapt their social media strategies to accommodate different linguistic, cultural, and disciplinary contexts. Accordingly, this study addresses the following three research questions (RQs):

- **RQ1**. To what extent are SCIE journals present on X, and CSCD journals present on WeChat?

- **RQ2**. How does the X uptake of SCIE journals and the WeChat uptake of CSCD journals vary across different disciplines?

- **RQ3**. To what extent are SCIE journals and CSCD journals active on X and WeChat, respectively, as measured by their posting frequency and the user engagement generated?

## 2. Data and Methods

*2.1. Bibliographic dataset of scientific journals*

This study begins by constructing two bibliographic datasets of scientific journals. The first dataset consists of journals indexed in the Science Citation Index-Expanded (SCIE, hereafter referred to as *SCIE journals*), and the second comprises journals indexed in the Chinese Science Citation Database (CSCD, hereafter referred to as *CSCD journals*).

The SCIE is a multidisciplinary citation index within the Web of Science Core Collection, covering journal papers dating back to 1900. As of 2024, it indexes over 9,500 journals and more than 61 million records across a broad range of scientific disciplines, primarily in the natural, medical, and engineering sciences.[2] In this study, we used the 2022 list of SCIE journals compiled by Nishikawa-Pacher [29], who manually verified whether each journal had a presence on X. After cleaning and deduplication, our final dataset contained 9,556 unique SCIE journals. We then retrieved bibliometric information for these journals from an in-house Web of Science database maintained by the Centre for Science and Technology Studies (CWTS) at

---
[2] More information about SCI-E can be found at: https://clarivate.com/products/scientific-and-academic-research/research-discovery-and-workflow-solutions/webofscience-platform/web-of-science-core-collection/science-citation-index-expanded/ (Accessed on October 7, 2024).

Leiden University (version dated March 2023), including the total number of published papers and the cumulative citations of these papers up to 2022.

The CSCD is a multidisciplinary citation index of Chinese scientific journal papers, covering more than 1,200 journals and 5.9 million records from 1989 to 2024.[3] It focuses on journals across a range of disciplines related to the natural, medical, and engineering sciences, and is therefore often regarded as the Chinese counterpart to the SCIE [47]. Developed by the Chinese Academy of Sciences, the CSCD is also hosted on the Web of Science as its first non-English product. Each Chinese journal paper indexed by the CSCD includes English-language bibliographic data, abstracts, and research area information, enabling searches in both Chinese and English. The CSCD has been employed in many studies to analyze the bibliometric characteristics of Chinese journals and to map the scientific knowledge in China [48–50]. In 2022, we downloaded the list of CSCD journals from the official website of the Chinese Academy of Sciences (http://sciencechina.cn/scichina2/index.jsp), yielding 1,262 unique CSCD journals. We collected bibliometric data for these journals, including the number of published papers and the total number of citations received by these papers up to 2022, from the China National Knowledge Infrastructure (CNKI, https://www.cnki.net/), which is the most comprehensive repository of academic and non-academic journals published in mainland China since 1915.[4]

The resulting lists of SCIE journals and CSCD journals form the foundation for our subsequent analyses, allowing us to examine the X uptake of internationally oriented English-language journals and the WeChat uptake of locally oriented Chinese-language journals, respectively.

*2.2. Disciplinary information of scientific journals*

To conduct the disciplinary analysis, we utilized the research area classification scheme available in the Web of Science database. This scheme applies to both SCIE and CSCD journals[5], thus enabling cross-comparison based on a standardized set of subject categories. Within this framework, journals are grouped into five broad *research area categories*, as outlined in Table 1.

---

[3] More information about CSCD can be found at: https://clarivate.com/products/scientific-and-academic-research/research-discovery-and-workflow-solutions/chinese-science-citation-database/ and https://webofscience.help.clarivate.com/Content/chinese-science/chinese-science-citation-database.htm (Accessed on October 7, 2024).

[4] More information about CNKI can be found at: https://www.cnki.net/gycnki/gycnki.htm (Accessed on October 20, 2024).

[5] More information about the classification of research areas in the Web of Science can be found at:

Table 1. Distribution of scientific journals across research area categories.

| Research area category | Number of research areas involved | Number of SCIE journals | Number of CSCD journals |
|---|---:|---:|---:|
| Life Sciences & Biomedicine | 75 | 5,534 | 526 |
| Technology | 21 | 1,941 | 486 |
| Physical Sciences | 17 | 1,837 | 227 |
| Social Sciences | 20 | 190 | 21 |
| Arts & Humanities | 8 | 54 | 2 |

For the SCIE journals, research area information was retrieved from the CWTS in-house Web of Science database; for the CSCD journals, the same information was manually collected from the Web of Science website. In cases where a journal was classified under multiple research areas, we applied fractional counting. Table 1 presents the total number of specific research areas within each broad category, as well as the number of SCIE and CSCD journals mapped to those categories. Overall, Life Sciences & Biomedicine constitutes the primary research area category for both SCIE and CSCD journals, followed by Technology and Physical Sciences. Because SCIE and CSCD journals predominantly focus on the natural, medical, and engineering sciences, relatively few interdisciplinary journals are classified under Social Sciences and Arts & Humanities. More detailed disciplinary distributions are provided in Figures A1 and A2 in Appendix A.

To further explore variations in social media uptake across disciplines, we focus on the three predominant categories (Life Sciences & Biomedicine, Technology, and Physical Sciences) and examine their uptake levels at the more granular research area level within each category. We use the following measure of *relative uptake*:

---

https://webofscience.help.clarivate.com/Content/research-areas.html?Highlight=research%20area (Accessed on October 20, 2024).

$$Relative\ uptake = \frac{P_{ra}}{P_{rac}}$$

where $P_{ra}$ denotes the proportion of SCIE (or CSCD) journals in a given *research area* that have X (or WeChat) accounts, and $P_{rac}$ represents the overall proportion of SCIE (or CSCD) journals with X (or WeChat) accounts in the broader *research area category*. A relative uptake value greater than 1 indicates that the uptake in a specific research area exceeds the average uptake for its parent category; conversely, a value less than 1 indicates below-average uptake. A value equals to 1 signifies that the research area's uptake level is on par with the broader category average.

*2.3. X data collection for SCIE journals*

To determine how many SCIE journals maintain X accounts, we employed a three-step process:

First, we began with the SCIE journal list compiled by Nishikawa-Pacher [29], which already included some manually verified X accounts, identified by searching for the full journal names on X.

Second, we used tweet IDs provided by Altmetric.com in November 2022 to collect detailed information on all tweets and corresponding X users in March 2023, leveraging the X API while it was still freely accessible. From this dataset of X mentions of scholarly publications, we matched the full names of SCIE journals against the profile names of X accounts to identify additional potential accounts operated by these journals.

Lastly, we merged the accounts identified in the previous two steps and performed manual verification to remove any mismatches, yielding the final dataset of SCIE journals with verified X accounts.

For each verified journal account, we extracted detailed metrics from our X database. This included the total number of tweets posted since the account's creation, encompassing both tweets referencing scholarly publications and those not. Specifically, for tweets referencing scholarly publications (i.e., those recorded by Altmetric.com with tweet IDs), we retrieved the total number of four types of user engagement associated with each journal account:

- Likes: Number of unique X users who liked a tweet;
- Retweets: Number of unique X users who reposted a tweet;
- Quotes: Number of quote tweets (i.e., reposts with additional comments);

- Replies: Number of replies a tweet received.

*2.4. WeChat data collection for CSCD journals*

To identify how many CSCD journals have established WeChat official accounts,[6] we manually searched each journal's full name using WeChat's built-in search engine. We then examined the account descriptions and posted articles to confirm whether an official account was indeed created and managed by the corresponding CSCD journal.

For those WeChat official accounts verified to be owned by CSCD journals, we further collected detailed metrics, including the total number of published WeChat articles and four types of user engagement aggregated across all posted WeChat articles:

- Reads: Number of unique WeChat users who clicked the link to view the full text of a WeChat article;

- Likes: Number of unique WeChat users who liked a WeChat article;

- Wows: Number of unique WeChat users who recommended a WeChat article to friends by clicking the "Wow" icon;

- Comments: Number of comments a WeChat article received.

**3. Results**

This section is organized into three parts. First, we report the overall uptake of X among SCIE journals and WeChat among CSCD journals. Second, we compare the uptake patterns across different research areas. Finally, we investigate user engagement metrics to understand how journals interact with audiences on these platforms.

*3.1. Overall uptake of X and WeChat by journals*

Figure 1 presents the distribution of SCIE journals with or without X accounts and CSCD journals with or without WeChat accounts. Overall, 22.7% of SCIE journals in our dataset maintain an X account, whereas 84.4% of CSCD journals have established a WeChat official account. This sizable difference indicates that Chinese scientific journals are considerably more inclined to engage with a local social media platform

---

[6] A WeChat official account is a public profile that individual or organizational users can create to gather and engage with followers. It serves as a communication channel for the account owner to publish content, send notifications, provide services, and chat with followers. See more information about WeChat official accounts at: https://mp.weixin.qq.com/?lang=en_US (Accessed on October 20, 2024).

(i.e., WeChat) compared to the tendency of SCIE journals to adopt X. Both SCIE journals with X accounts and CSCD journals with WeChat accounts tend to receive more citations overall than those without such accounts (see Figure B1 in Appendix B), suggesting a potential relationship between a journal's academic impact and its social media presence.

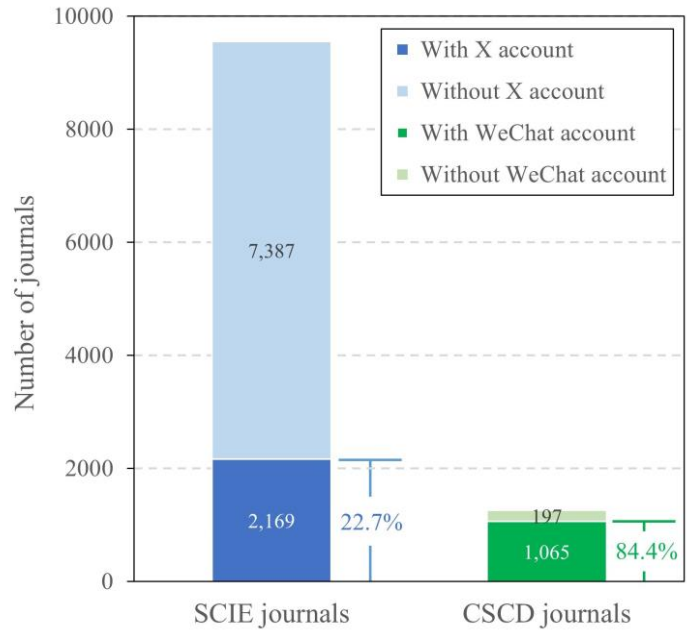

**Figure 1**. Distribution of SCIE/CSCD journals with or without X/WeChat accounts.

*3.2. Disciplinary comparison between the X and WeChat uptake of journals*

To compare how disciplinary factors may influence journals' decisions to adopt social media, we utilized the research area information from Clarivate. Figure 2 illustrates the proportion of SCIE journals with X accounts and the proportion of CSCD journals with WeChat accounts across five broad research area categories. Because the number of journals in the Social Sciences and Arts & Humanities categories is small in both SCIE and CSCD, they are less representative within our dataset. Among the remaining three categories, which are highlighted in darker colors, Life Sciences & Biomedicine exhibits the highest social media presence in both datasets, with 29.0% of SCIE journals on X and 85.7% of CSCD journals on WeChat. In contrast, SCIE journals in Technology and Physical Sciences display lower uptake on X – 12.6% and 13.7%, respectively – whereas CSCD journals in these same categories demonstrate relatively high WeChat

uptake – 85.2% for Technology and 79.7% for Physical Sciences.

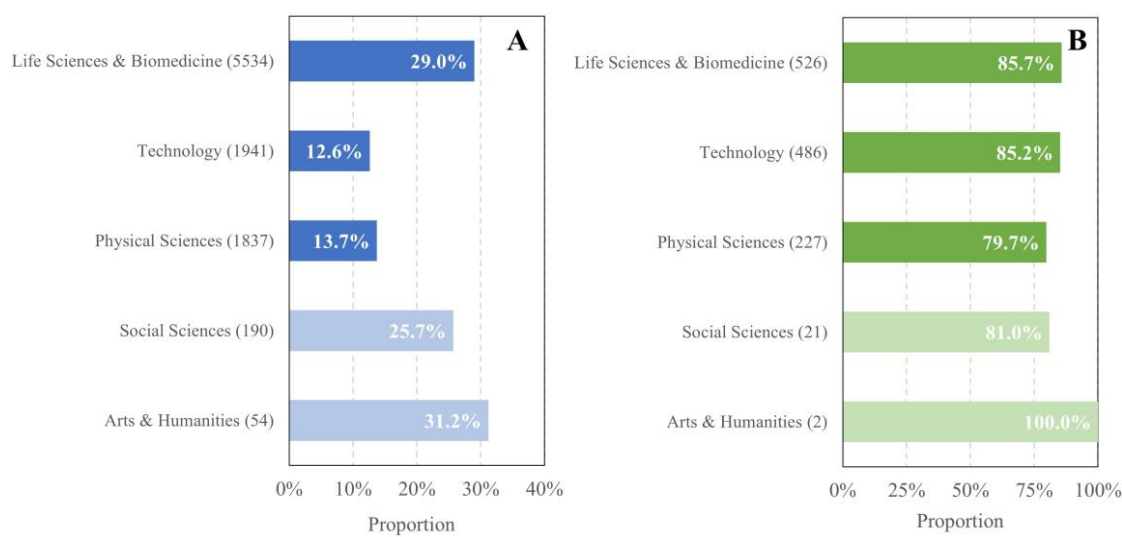

**Figure 2**. (A) X uptake of SCIE journals and (B) WeChat uptake of CSCD journals across five broad research area categories.

Figures 3 through 5 provide a more granular view of social media uptake within the three predominant research area categories – Life Sciences & Biomedicine, Technology, and Physical Sciences – by illustrating each specific research area's relative uptake on X (SCIE) and WeChat (CSCD). In the SCIE dataset (X), the variation across research areas is particularly pronounced, likely reflecting the wider global scope of research areas covered by SCIE journals. For CSCD (WeChat), the overall adoption rates are high, so differences in relative uptake are somewhat narrower. Nevertheless, notable discrepancies emerge when we compare the research areas with the highest and lowest relative uptake between X and WeChat. In many instances, the research areas that lead in X uptake among SCIE journals are not the same as those that excel in WeChat uptake among CSCD journals, and vice versa. This suggests that, at a more fine-grained level, SCIE and CSCD journals in different research areas show varying tendencies to create social media accounts, revealing disciplinary variations not only between social media platforms but also within a single platform.

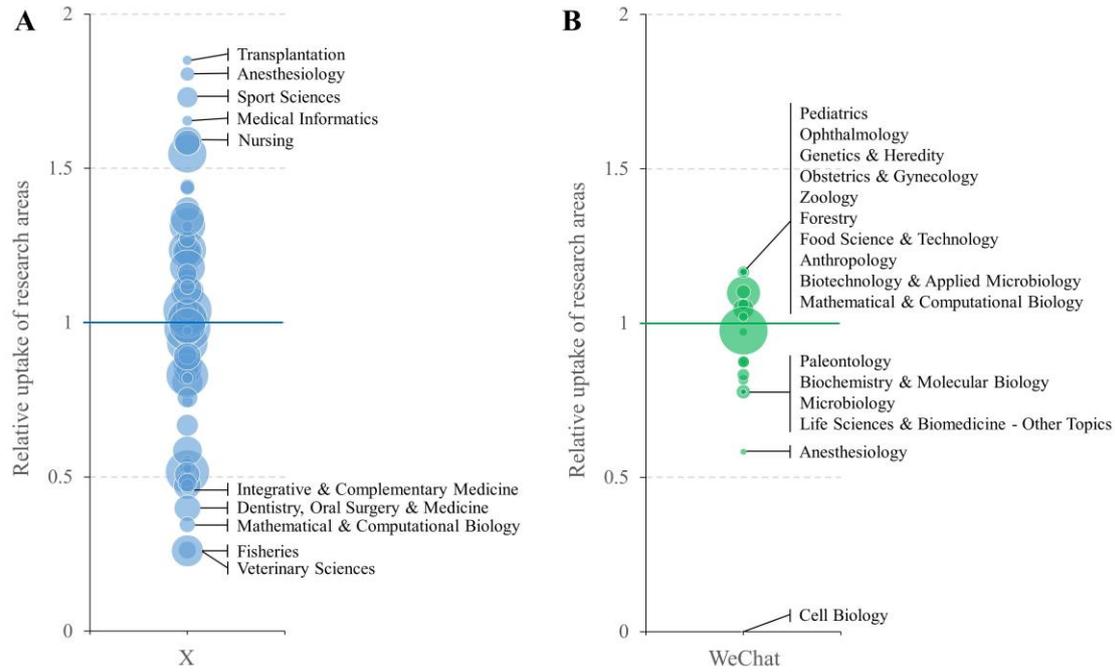

**Figure 3**. (A) Relative X uptake of SCIE journals and (B) relative WeChat uptake of CSCD journals in Life Science & Biomedicine. Bubble size indicates the number of journals in each research area. The five research areas with the highest and lowest relative uptake are annotated. Note that ten research areas share the highest relative uptake value on WeChat and are therefore listed collectively.

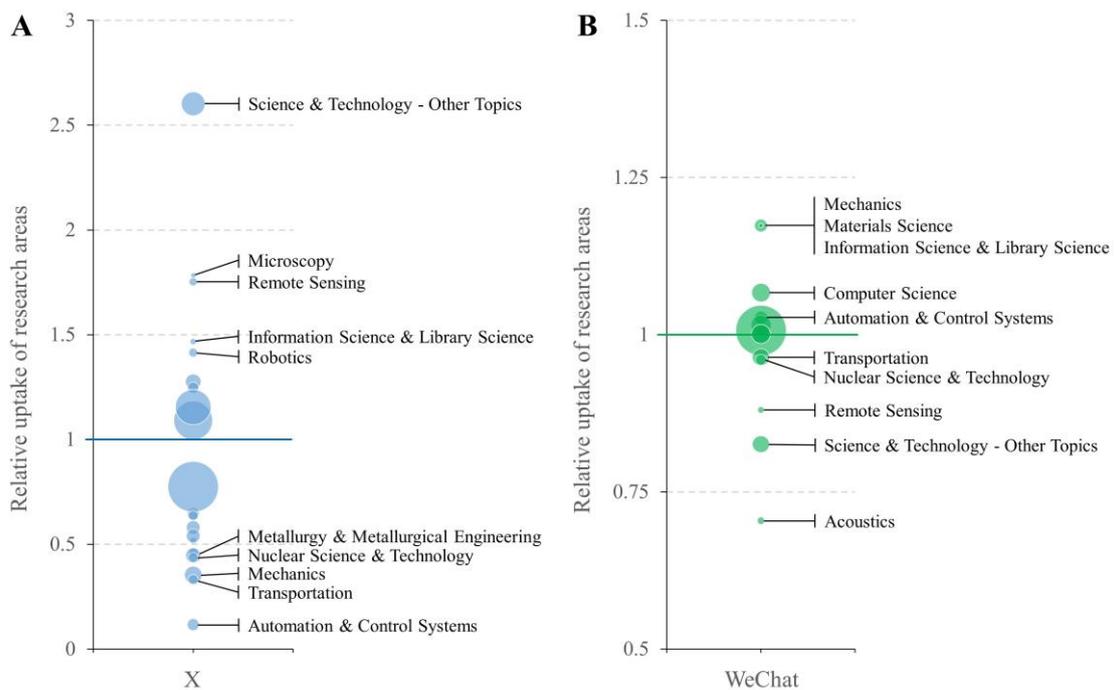

**Figure 4**. (A) Relative X uptake of SCIE journals and (B) relative WeChat uptake of CSCD journals in Technology. Bubble size indicates the number of journals in each research area. The five research areas with the highest and lowest relative uptake are annotated.

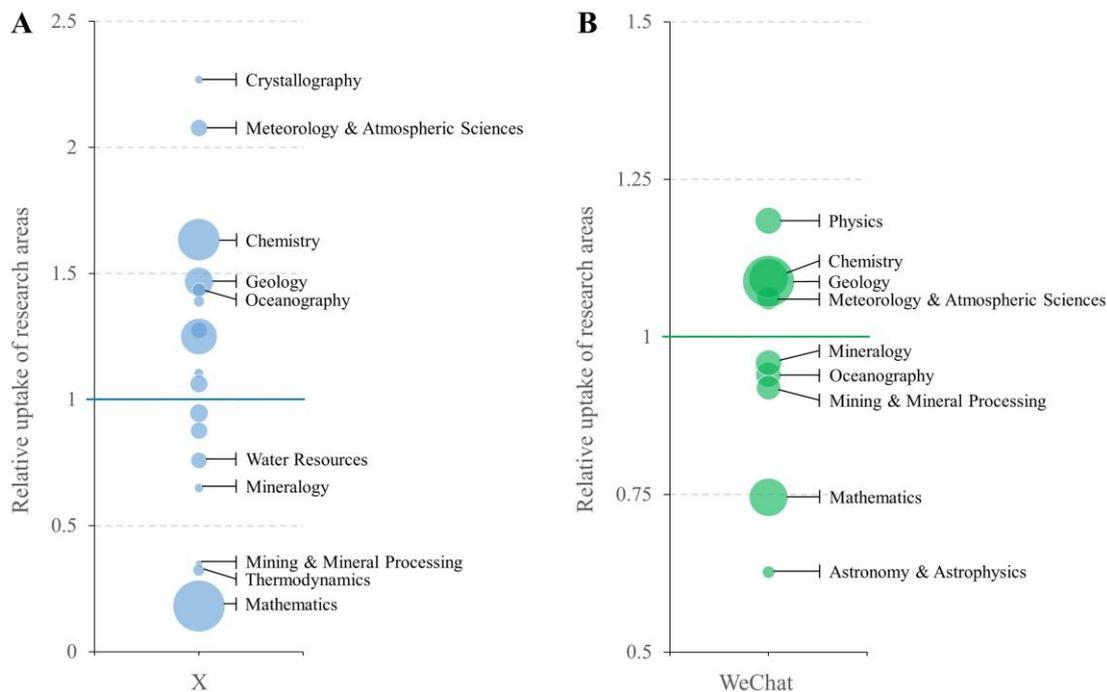

**Figure 5**. (A) Relative X uptake of SCIE journals and (B) relative WeChat uptake of CSCD journals in Physical Sciences. Bubble size indicates the number of journals in each research area. The five research areas with the highest and lowest relative uptake are annotated. Note that only four research areas have relative uptake values above 1 on WeChat, so the top four, rather than five, research areas are listed.

*3.3. Comparison of posting frequency and user engagement of journals on X and WeChat*

To evaluate how actively SCIE and CSCD journals use X and WeChat, respectively, we first calculated the ratio of social media posts to scientific papers published for each journal. Specifically, for an SCIE journal, this ratio was computed as the total number of tweets it posted divided by the total number of papers it published; likewise, for a CSCD journal, it was the number of WeChat articles posted divided by the total number of papers published. A value greater than 1 indicates that the journal posts more social media content than it publishes scientific papers, whereas a value less than 1 suggests the opposite.

As shown in Figure 6, both SCIE and CSCD journals generally publish more papers than they post social media content, although this tendency is more pronounced among CSCD journals on WeChat. Only 4% of CSCD journals had a ratio above 1 (posting more WeChat articles than publishing papers), while over 21% of SCIE journals showed a higher volume of tweets than scientific publications. These results suggest that, among journals that maintain social media accounts, SCIE journals on X tend to generate social media content more frequently compared to CSCD journals on WeChat.

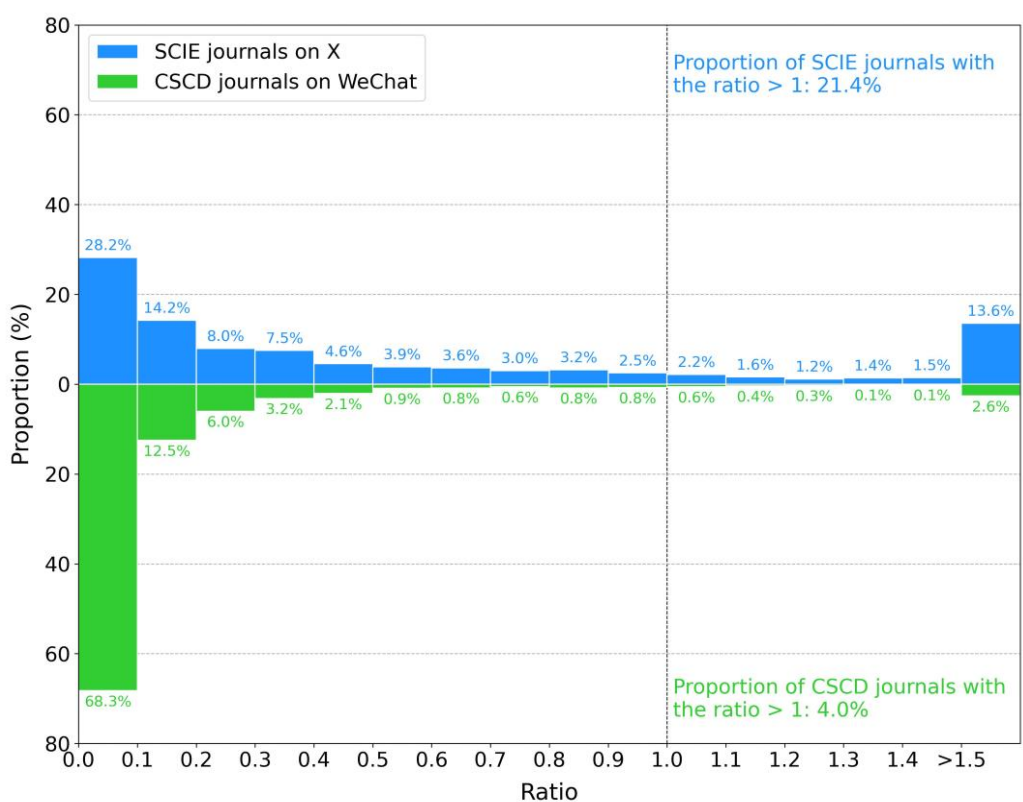

**Figure 6**. Distritbution of SCIE journals on X and CSCD journals on WeChat according to the ratio of social media posts to published papers.

To further examine how actively users engage with journal-operated social media accounts, Figure 7 shows the distribution of five types of X metrics for SCIE journals and five types of WeChat metrics for CSCD journals. Each boxplot exhibits substantial variations, indicating that journals differ markedly in their ability to stimulate user engagement. SCIE journals generally posted more tweets in total than CSCD journals posted WeChat articles, aligning with the observations in Figure 6. This disparity may reflect differences in platform norms: tweets usually comprise concise text and links, whereas WeChat articles often involve longer, more detailed content, leading to lower posting frequency.

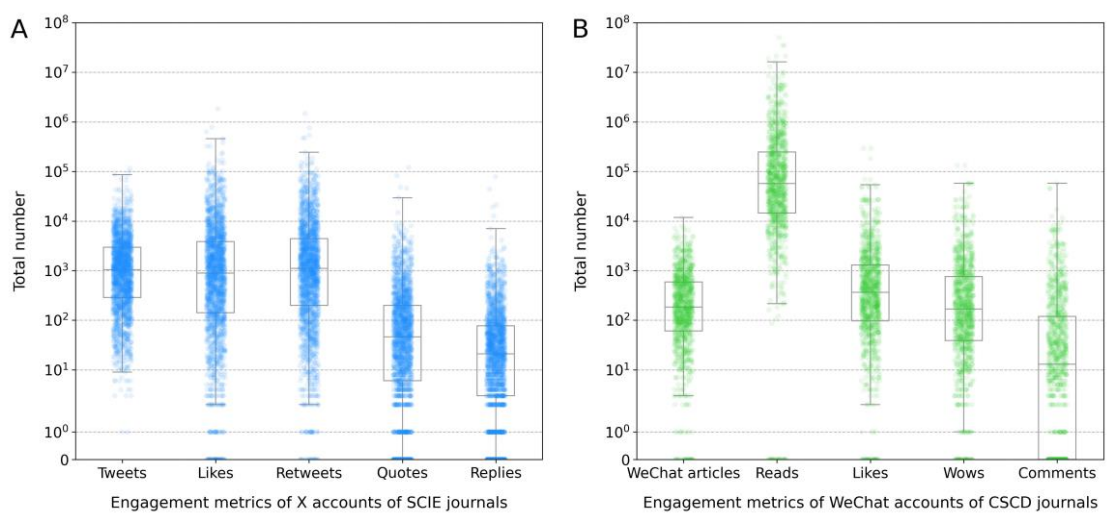

**Figure 7**. Distribution of (A) X metrics of SCIE journals and (B) WeChat metrics of CSCD journals.

On X, "likes" and "retweets" are the most common engagement types, requiring minimal user effort (e.g., clicking a "like" or "retweet" button). "Quotes" and "replies", which require additional user-generated content, are less frequent. On WeChat, "reads" appear most frequently because simply clicking on an article to view it is a low-effort action. "Likes" and "wows" are also relatively common; both can be performed by clicking an icon. "Comments" are comparatively rarer, similar to quotes and replies on X, because they necessitate additional content creation. Furthermore, WeChat comments must be approved by the account owner before being displayed, which adds another hurdle to user engagement.

Figures 8 and 9 present Spearman correlation analyses between bibliometric indicators (i.e., number of published papers and total citations) and social media metrics (i.e., X metrics for SCIE journals and WeChat metrics for CSCD journals). Both X and WeChat show weak-to-moderate correlations between bibliometric indicators and social media metrics. However, these relationships are notably weaker for CSCD journals on WeChat, suggesting more pronounced discrepancies between bibliometric indicators and social media metrics in the Chinese context.

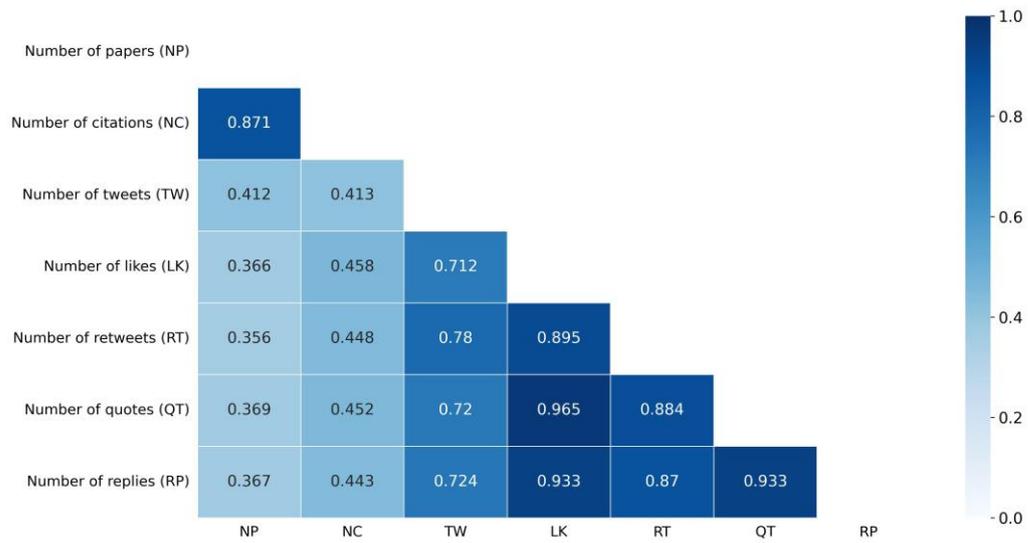

**Figure 8**. Spearman correlations among bibliometric and X metric indicators of SCIE journals.

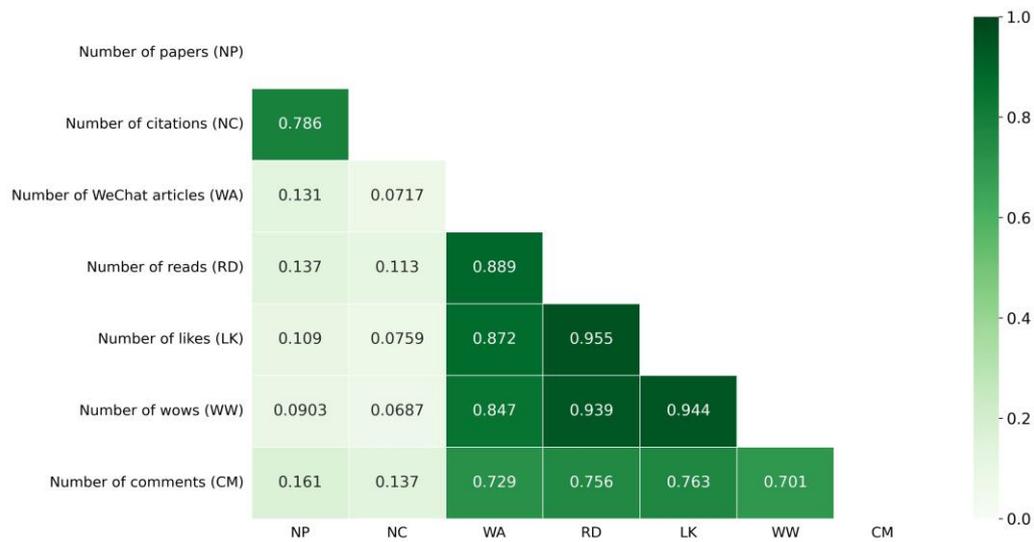

**Figure 9**. Spearman correlations among bibliometric and WeChat metric indicators of CSCD journals.

## 4. Discussion

This study compared how scientific journals indexed in the SCIE and the CSCD utilize two major social media platforms – X and WeChat, respectively – and how their audiences engage with the content shared on these platforms. By juxtaposing international journals on a global platform (X) with local journals on a domestic platform (WeChat), this study broadens the scope of social media metrics and underscores the importance of examining science communication in more localized contexts.

*4.1. Different levels of X and WeChat uptake among scientific journals*

A central finding of this study is the pronounced discrepancy between the X uptake of SCIE journals (22.7%) and the WeChat uptake of CSCD journals (84.4%). These divergent uptake rates highlight the role of local contexts (international vs. Chinese) and platform popularity (X vs. WeChat) in shaping the decisions of a journal to establish a social media presence. WeChat has become deeply integrated into China's digital ecosystem, offering a comprehensive range of features (e.g., instant messaging, microblogging, and mobile payments) that make it a dominant platform for both personal and institutional communication [51–53]. Consequently, Chinese scientific journals find it both strategically and culturally imperative to maintain a WeChat official account to reach and engage diverse domestic audiences.

In contrast, although the proportion of SCIE journals with an X account has risen somewhat compared to earlier findings [30], it remains relatively low overall. Several factors could account for this. First, many English-language scientific journals rely on publisher-level X accounts for broader marketing rather than creating separate journal-level accounts. Second, X is not the only international platform available: other sites, such as Facebook, and more recently Bluesky or Mastodon, may be prioritized. This is evidenced by higher Facebook uptake in certain fields and greater user engagement compared to X [30, 37, 40]. Additionally, the instability of global social media platforms, such as Twitter's rebranding to X and substantial changes to its communication and engagement policies (e.g., reduced fact-checking and moderation tools) [54–56], may contribute to hesitancy among journals to invest heavily in the platform. Finally, some journals may lack resources or personnel dedicated to consistent social media outreach, which can hinder their presence on social media platforms.

At the disciplinary level, Life Sciences & Biomedicine exhibits the highest social media uptake on both platforms, likely reflecting heightened public interest in medical and biological research, as reported by previous research at the paper level [57, 58].

Technology and Physical Sciences show lower uptake rates on X but remain relatively active on WeChat, a discrepancy suggesting that on WeChat, local audience expectations and the platform's multifunctionality create incentives for journals – even those in more specialized and natural scientific domains – to establish official accounts as a matter of cultural and professional expectation within China. Given the integral role of WeChat in Chinese scholarly publishing [59, 60], incorporating this platform into the realm of social media metrics is critical, particularly for research with local relevance in China.

*4.2. Similarities and differences between X and WeChat metrics*

Our exploration of user engagement metrics reveals both commonalities and distinctions between X and WeChat. SCIE journals on X generally produce posts more frequently, whereas CSCD journals on WeChat generate relatively fewer articles. This difference aligns with the platforms' inherent characteristics: X's concise format facilitates rapid updates and real-time discussions, whereas WeChat's official accounts lend themselves to longer-form articles akin to mini "newsletter" posts or extended commentaries.

On both platforms, lower-effort forms of engagement (e.g., "likes", "retweets", "reads", and "wows") generally outnumber higher-effort interactions (e.g., "quotes", "replies", and "comments"), consistent with the observation that users engage more often with quick and easily executed forms of reaction [61]. Consequently, while scientific journals often achieve basic forms of user engagement, fostering more in-depth dialogue on social media remains challenging, whether on an international or a local platform.

Finally, the correlation analyses show that bibliometric indicators at the journal level bear only weak-to-moderate correlations with social media engagement metrics, echoing similar findings at the paper level [57, 61, 62]. In other words, the academic impact of a journal – whether measured by publications or citations – does not necessarily translate into high levels of social media engagement. This pattern holds for both X and WeChat, but is more pronounced on WeChat, suggesting that WeChat-based interactions may be driven less by a journal's scholarly reputation and more by factors like engaging content and marketing strategies. Since content analysis offers deeper insights into the mechanisms behind how and why social media metrics are generated [63], future research could benefit from closely examining the nature of WeChat articles posted by journals. Such research may clarify how local platform characteristics and culturally specific practices shape visibility and engagement, and how WeChat can be effectively integrated into broader social media studies of science.

*4.3. Limitations and future directions*

This study has several limitations that warrant discussion. First, although X and WeChat are both prominent platforms for the dissemination of scientific information, they differ considerably in terms of user base, functionality, and interaction design. As such, they are not strictly equivalent. However, they share fundamental characteristics of social media platforms: they are Web 2.0-based, enable user-generated content, and support interactive communication. More importantly, each plays a dominant role in its respective sociolinguistic context – X in the global, predominantly English-speaking academic environment, and WeChat within the Chinese-speaking scholarly community. This contextual predominance was a key rationale for selecting these platforms as the basis for comparison in our study. Nonetheless, we acknowledge that other platforms with more similar functional architectures – such as X and Weibo, or Instagram and RedNote – also have substantial user bases and may serve important roles in science communication. Journals may adopt different platforms depending on their outreach strategies, target audiences, or disciplinary norms, potentially leading to different patterns of social media presence and engagement. Future research should expand the comparative scope to include a wider range of platforms, particularly those that are structurally more comparable, to develop a more comprehensive understanding of social media usage in science communication across diverse contexts.

Second, our analysis focuses primarily on the presence and overall engagement levels of scientific journals on social media platforms, without examining the content of the posts themselves. As noted earlier, content analysis can offer critical insights into the communication strategies and motivations underlying journals' social media activity. Moreover, the topical focus, tone, and format of content may significantly influence the level of attention and interaction it receives [64]. Understanding what types of content generate engagement is therefore essential for interpreting social media metric indicators and for evaluating the communicative effectiveness of journal-led social media efforts. Future research should incorporate detailed content analyses, where data access permits, to better understand the factors driving social media attention and the strategic choices journals make in their digital communication.

Third, our study does not include an analysis of the users who engage with journal accounts. The demographics and identities of these users – whether they are researchers, students, journalists, or members of the general public – as well as the nature of their interactions (e.g., passive sharing vs. active discussion), are crucial for understanding the broader impact of journal communication on social media. Due to data limitations, we were unable to retrieve comprehensive user-level information from either X or WeChat. However, we recognize this as an important limitation and highlight it as a key direction for future work. Future studies should aim to integrate user-level data to offer a more nuanced understanding of audience engagement. Such research would deepen our knowledge of the user-content-motivation dynamics and help inform more effective science communication strategies via social media platforms.

Fourth, a notable methodological limitation of this study lies in the difference in how engagement metrics were derived for the two platforms. For X, engagement metrics were based solely on tweets referencing scholarly publications, as identified via Altmetric.com, due to limitations in data availability. In contrast, the WeChat metrics reflect engagement with all WeChat articles posted by journal accounts, regardless of whether they referenced scientific publications. This discrepancy in the scope of content included may affect the comparability of engagement levels across platforms. To address this limitation and avoid misleading conclusions, we deliberately restricted our analysis to within-platform comparisons of engagement types, rather than performing direct comparisons between X and WeChat. Future research could benefit from more equivalent datasets that include comprehensive post-level data from both platforms to enable more balanced and robust cross-platform engagement analyses.

Fifth, to accurately identify the social media accounts of scientific journals on X and WeChat, we undertook extensive manual verification efforts. This included checking the validity of X accounts identified through existing open datasets and our own matching process, as well as scrutinizing WeChat accounts retrieved via name-based searches. While we made every effort to ensure the accuracy of these identifications, it is not possible to completely eliminate the risk of mismatches or omissions. Potential issues may arise due to changes in journal titles, the renaming of social media accounts, or inconsistencies between the official journal names and their customized social media profile names. Nevertheless, we have clearly documented the identification methodology employed, which can be replicated and refined in future research to validate and update journal uptake data on X and WeChat.

Lastly, it is important to acknowledge that the rapidly evolving nature of social media platforms introduces additional complexity to this line of research. For instance, some scientific journals may have created new social media accounts, or additional user engagement may have occurred after the period of data collection, potentially leading to changes in the patterns observed in this study. As such, our findings should be interpreted as a snapshot reflecting the specific time frame analyzed. Future research would benefit from longitudinal analyses that track the development of journals' social media adoption and engagement over time. Moreover, the shifting popularity and functionality of platforms further highlight the need for continuous monitoring and adaptation in the study of social media use in science communication. For instance, Twitter's rebranding to X and changes to its content moderation policies could impact user engagement and the platform's overall effectiveness for science communication. This shift has already been reflected in the growing migration of academics and scientific institutions from X to alternative platforms such as Bluesky and Mastodon [56, 65]. Similarly, Facebook's recent removal of certain moderation features may alter user interactions with scientific content [66]. Potential future changes to WeChat, which currently dominates in China, could also affect how scientific journals engage with their audiences. As highlighted by Haustein [67], these evolving dynamics may pose significant challenges to social media metrics, underscoring the need for continuous adaptation in measuring and interpreting social media activities across platforms – particularly for promising platforms like Bluesky and WeChat in the "post-Twitter" era.

## 5. Conclusions

This study presents a comparative examination of the X uptake of SCIE journals and the WeChat uptake of CSCD journals, revealing notable discrepancies in platform adoption and user engagement. While only 22.7% of SCIE journals have an X account, 84.4% of CSCD journals maintain a WeChat official account. Journals in Life Sciences & Biomedicine demonstrate the highest uptake on both platforms, whereas those in Technology and Physical Sciences show comparatively lower uptake on X but a robust presence on WeChat. In terms of posting frequency, SCIE journals on X tend to generate social media content more frequently compared to CSCD journals on WeChat. Regarding user engagement, both X and WeChat are characterized by higher frequencies of low-effort interactions (e.g., "likes" and "reads") compared to more interactive engagements (e.g., "quotes" and "comments"). Correlation analyses indicate weak-to-moderate correlations between bibliometric indicators and social media metrics, confirming that online engagement may capture a distinct facet of journal impact, whether on an international or a local platform.

These findings underscore consistent themes across social media contexts – such as disciplinary variations, platform variations, limited conversational interactions, and modest correlations with bibliometric indicators – yet they also highlight the prominent roles that local platforms can play in the visibility of regional research outputs. Accordingly, expanding social media metrics to incorporate additional locally relevant platforms would enrich our understanding of science communication and better capture the diverse ways in which scientific information is disseminated and engaged with worldwide.

**Acknowledgments**

The authors thank Altmetric.com for providing the data for research purposes, Dangqiang Ye for assisting with the WeChat data collection, and the anonymous reviewers for their valuable suggestions.

**Statements and Declarations**

Not applicable

**Declaration of conflicting interest**

The author(s) declared no potential conflicts of interest with respect to the research, authorship, and/or publication of this article.

**Funding statement**

The author(s) disclosed receipt of the following financial support for the research, authorship and/or publication of this article: This study is funded by the National Natural Science Foundation of China (No. 72304274) and the National Social Science Foundation of China (No. 24FXWB004). Er-Te Zheng is financially supported by the GTA scholarship from the School of Information, Journalism and Communication of the University of Sheffield. Rodrigo Costas is partially funded by the South African DSI-NRF Centre of Excellence in Scientometrics and Science, Technology and Innovation Policy (SciSTIP).

**Appendix A**

The tree maps in Figures A1 and A2 illustrate the distribution of SCIE journals and CSCD journals across various research areas, respectively. Research areas that share the same color belong to the same broad category. Although CSCD journals span fewer research areas than SCIE journals, they exhibit a similar overall pattern: Life Sciences & Biomedicine comprises the largest share, followed by Technology and Physical Sciences, while Social Sciences and Arts & Humanities account for only a small fraction. This distribution reflects the specific emphasis of both databases on the natural, medical, and engineering sciences.

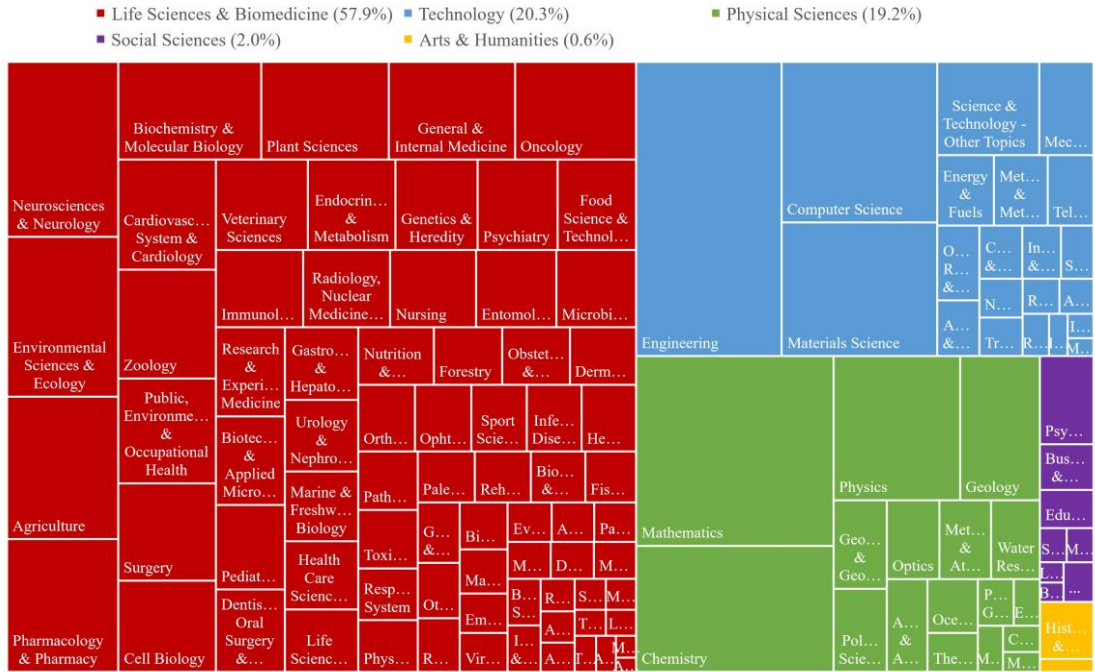

**Figure A1**. Distribution of SCIE journals across research areas.

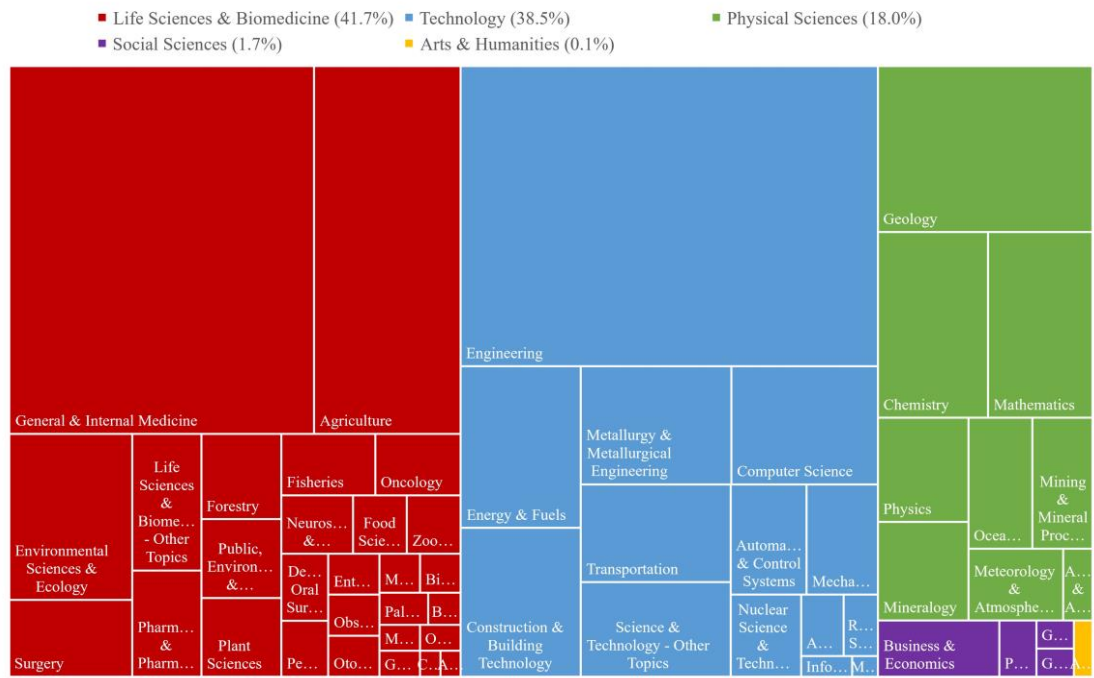

**Figure A2**. Distribution of CSCD journals across research areas.

# Appendix B

Figure B1 compares the total number of citations (log-transformed) received by journals with and without a presence on X and WeChat. The figure shows that journals with a presence on either platform tend to have significantly higher citation counts than those without. This citation advantage is particularly pronounced for journals with a presence on X. The differences were confirmed to be statistically significant using the Mann-Whitney U test (X: $p < 0.001$; WeChat: $p < 0.001$).

These findings suggest a potential association between social media uptake and academic impact. However, whether social media presence drives increased citations – or whether more prominent journals are simply more likely to adopt social media – remains an open question. Future research should further explore this relationship to better understand the role of social media in shaping the visibility and strategic development of scientific journals.

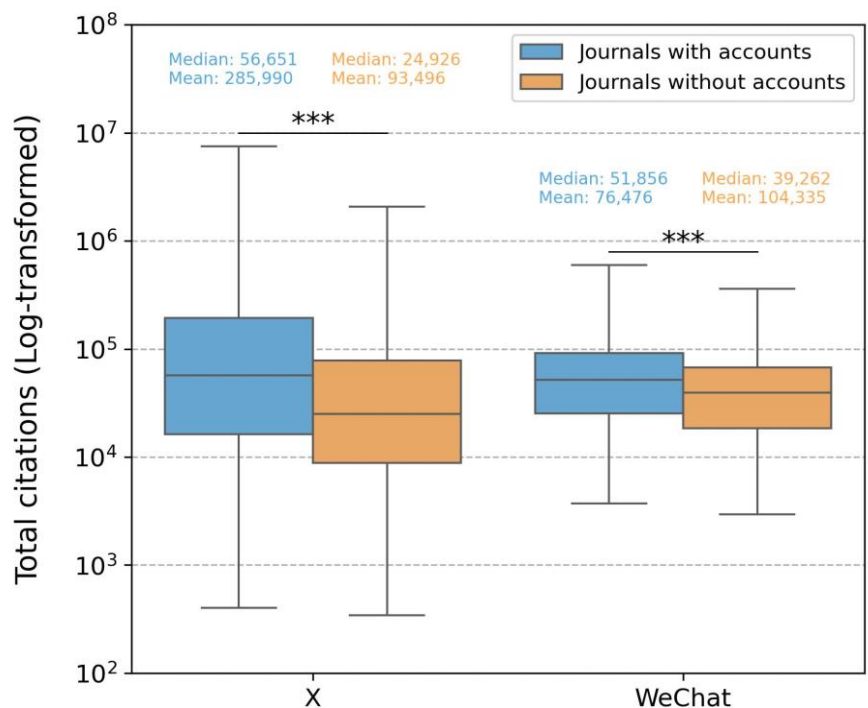

**Figure B1**. Comparison of total citations received by journals with or without a presence on X and WeChat. *** indicates statistical significance at $p < 0.001$.